\documentclass[fleqn,usenatbib]{mnras}

\usepackage{newtxtext,newtxmath}

\usepackage[T1]{fontenc}
\usepackage{ae,aecompl}
\usepackage{color}

\usepackage{graphicx}	
\usepackage{amsmath}	
\usepackage{amssymb}	

\title[UFO in the TDE ASASSN-14li]{Ultrafast Outflow in Tidal Disruption Event ASASSN-14li}

\author[E. Kara et al.]{
E. Kara,$^{1,2,3}$\thanks{E-mail: ekara@astro.umd.edu}, L. Dai$^{4,5,2}$, C. S. Reynolds$^{6,1,2}$ and T. Kallman$^{3}$
\\
$^{1}$Department of Astronomy, University of Maryland, College Park, MD 20742\\
$^{2}$Joint Space Science Institute, University of Maryland, College Park, MD, 20742\\
$^{3}$X-ray Astrophysics Laboratory, NASA/Goddard Space Flight Center, Greenbelt, MD 20771\\
$^{4}$Dark Cosmology Centre, Niels Bohr Institute, Juliane Maries Vej 30, 2100 Copenhagen, Denmark\\
$^{5}$Department of Physics, University of Maryland, College Park, MD, 20742\\
$^{6}$Institute of Astronomy, Madingley Rd, Cambridge CB3 0HA, United Kingdom\\
}

\date{Accepted XXX. Received 13 November 2017; in original form 12 October 2017}

\pubyear{2017}

\begin{document}
\label{firstpage}
\pagerange{\pageref{firstpage}--\pageref{lastpage}}
\maketitle

\begin{abstract}
At only 90~Mpc, ASASSN-14li is one of the nearest tidal disruption event (TDE) ever discovered, and because of this, it has been observed by several observatories at many wavelengths. In this paper, we present new results on archival {\em XMM-Newton} observations, three of which were taken at early times (within 40 days of the discovery), and three of which were taken at late times, about one year after the peak. We find that, at early times, in addition to the $\sim10^5$~K blackbody component that dominates the X-ray band, there is evidence for a broad, P~Cygni-like absorption feature at around 0.7~keV in all {\em XMM-Newton} instruments (CCD detectors and grating spectrometers), and that this feature disappears (or at least diminishes) in the late-time observations.  We perform photoionization modelling with {\sc XSTAR} and interpret this absorption feature as blueshifted OVIII, from an ionized outflow with a velocity of 0.2c. As the TDE transitions from high to low accretion rate, the outflow turns off, thus explaining the absence of the absorption feature during the late-time observations.
\end{abstract}




\section{Introduction}
Roughly once every hundred thousand years, a supermassive black hole (SMBH) at the center of a galaxy tidally disrupts an orbiting star \citep{rees88,phinney89}. About half of the stellar debris continues on its original trajectory, while the other half becomes gravitationally bound and falls back towards the supermassive black hole. The accretion of this bound stellar material causes a short-lived flare of emission, known as a tidal disruption event (TDE). If the central SMBH is less than $10^{7} M_{\odot}$, the fallback rate can initially exceed the Eddington limit and then decays over time with a well-known $t^{-5/3}$ dependence. If the bound stellar debris can lose angular momentum and circularize quickly \citep{dai15}, then the resulting accretion onto the SMBH (and the luminosity) may also obey the same time dependence. Observations of such events have been seen in the optical (e.g. \citealt{gezari07}), soft X-rays (e.g. \citealt{komossa15}), and even sometimes in hard X-rays (e.g. \citealt{burrows11, cenko12}). 


The prediction of super-Eddington fall-back rates, and potentially super-Eddington accretion rates if circularization is efficient (e.g. \citealt{guillochon14}, \citealt{dai15}) make TDEs an exciting case study for understanding how rapidly black holes can grow (e.g. \citealt{lin17}), which may play a particularly important role in the formation of the earliest supermassive black holes. Several recent theoretical efforts have been made in understanding super-Eddington accretion in TDEs (e.g. \citealt{coughlin14,sadowski16,mckinney15}). Although the overall radiative efficiency and luminosity are still debated, in all simulations the disk structure deviates greatly from a standard thin disc, and instead there is geometrically and optically thick accretion flow with strong, fast outflows.

Ultrafast outflows (UFOs) are defined as massive outflows with velocity greater than 10,000~km/s. 
They have been seen in a number of active galactic nuclei (e.g. \citealt{pounds03,tombesi13,parker17}), in Galactic Black Hole Binaries in the high soft state \citep{king12}, in ultraluminous X-ray sources \citep{pinto16,pinto17}, and in a few TDEs \citep{lin15,kara16}. They are predicted in super-Eddington accretion flows (e.g. \citealt{laor14,cmiller15}), but also can be driven in sub-Eddington systems as well (e.g. \citealt{fukumura15}). In this paper, we present evidence for an ultrafast outflow in the TDE, ASASSN-14li.


\begin{figure}
\includegraphics[width=\linewidth]{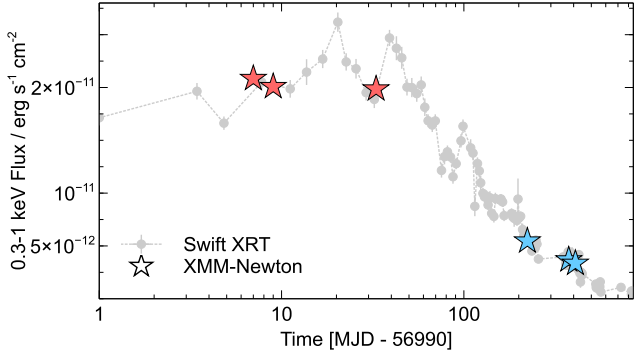}\\
\caption{The longterm {\em Swift}-XRT light curve of ASASSN-14li (gray points), overlayed with the {\em XMM-Newton} pointed observations shown in filled stars. The `early-time' observations are shown in red, and the `late-time' observations in blue.}
\label{lc}
\end{figure}

\begin{table}
\begin{tabular}{c c c c c}
\hline
\vspace{0.1cm}
{\bf (a)} & {\bf (b)} & {\bf (c)} & {\bf (d)} & {\bf (e)} \\
{\bf obsid} & {\bf Date} & {\bf Length/Exp.} & {\bf Flux} & {\bf Reg.} \\
\hline
0694651201	&	2014.12.06	&	21470/9380 &	2.1 & 10 \\
0722480201	&	2014.12.08	&	93470/21320 &	2.0 & 14 \\
0694651401	&	2015.01.01	&	23270/16320 & 2.0 & 13 \\
0694651501	&	2015.07.10	&	21970/15010 & 0.55 & 7 \\
0770980101	&	2015.12.10	&	94970/55050 & 0.37 & 9 \\
0770980501	&	2016.01.12	&	8047/4192 & 0.35 & 0 \\
\hline
\end{tabular}
\caption{{\em XMM-Newton} observations used in this analysis. The columns are: (a) Observation ID, (b) observation start date, (c) total observation length and final exposure in seconds, (d) flux measured from 0.3-1 keV in units of $10^{-11}$~erg~cm$^{-2}$~s$^{-1}$, (e) radius of the inner circular region excised to make an annular region to mitigate pile-up (see Appendix).}
\label{obs_table}
\end{table}


ASASSN-14li \citep{jose14} was discovered by the  All-Sky Automated Survey for SuperNovae (ASAS-SN; \citealt{shappee14}) on 2014-11-22 (MJD = 56983.6).  It was followed up by a Swift campaign \citep{holoien16}, revealing strong X-ray emission that led to Target of Opportunity observations with {\em XMM-Newton} and {\em Chandra} that revealed a highly-ionised, low-velocity outflow in the grating spectra \citep{miller15}. This low-velocity outflow was also seen in lower ionization species, through a dedicated {\em Hubble Space Telescope} observation \citep{cenko16}. In addition to optical, X-ray and UV observations, the source was followed-up in the radio band by two groups. The radio emission has been interpreted in two very different ways: as a wide-angle outflow of 12,000-36,0000 km/s \citep{alexander16}, or as a compact jet \citep{vanvelzen16,pasham17}.  An IR dust echo was detected $\sim 100$~days after the peak emission, suggesting irradiation from a central source of $L=10^{43}-10^{45}$~erg~s$^{-1}$, consistent with the bolometric luminosity from UV/X-ray observations \citep{jiang16}.

 
\section{Observations and Data Reduction}
\label{obs}
 
\begin{figure}
\includegraphics[width=\columnwidth]{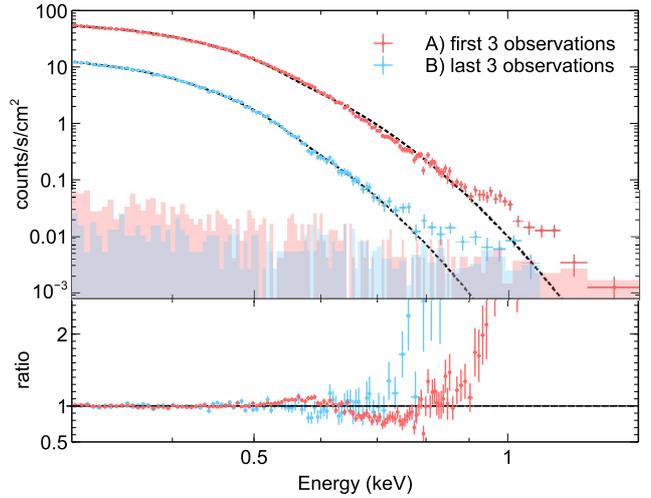}\\
\caption{The average early-time (red) and late-time (blue) spectra, and their corresponding backgrounds. The best fit blackbody for both spectra are shown as dotted black lines. The bottom panel shows the ratio of the spectra to their best fit blackbody models. The ratio shows a hard excess, which is more dominant in the late-time spectrum.}
\label{spec}
\end{figure}



In this letter, we focus on six {\em XMM-Newton} observations of ASASSN-14li (see Table~\ref{obs_table} and Fig.~\ref{lc}).  
We use the data from the {\em XMM-Newton} EPIC-pn camera \citep{struder01,jansen01}. These data were not included in the original {\em XMM-Newton} paper by \citet{miller15}.  We reduced the data using the {\em XMM-Newton} Science Analysis System (SAS v. 14.0.0) and the newest calibration files. We started with the observation data files and followed standard procedures. The source extraction regions are circular regions of radius 35 arcsec centered on the maximum source emission.  The background regions are also circular regions of radius 35 arcsecs.

Several of the observations were strongly affected by background flares. 
We used a very conservative background rate cut in order to produce the highest signal-to-noise spectrum at high energies as possible. The good-time-intervals were selected to be intervals where the count rate in the 12-15 keV light curve was less than 0.1 counts/s. The resulting source and background-subtracted spectra are identical within the error bars because the resulting background is sufficiently low. 
The spectrum becomes background dominated at 1.5~keV for the first three observations, and at 1.1~keV for the last three observations. Therefore, we use these energies as the respective upper bounds of the spectral analysis.

All six observations were taken in Small Window Mode because the source was very bright, especially at early times.  Though the count rate is nominally below the Small Window Mode threshold for pile-up, the counts are concentrated at low energies in this very soft source, and so pile-up was an issue in the CCD detectors. To mitigate pile-up effects, we used only single events spectra. See Appendix for details on how we deal with pile-up.

The response matrices were produced using rmfgen and arfgen in SAS. The pn spectra were binned to a minimum of 25 counts per bin.

The RGS data were reduced using the standard pipeline tool, {\sc rgsproc}, which produces the source, background spectral files, and instrument response files. We binned the RGS 1 and 2 spectra to 1 count per bin and fit them jointly using the C-statistic.

Spectral fitting was performed using xspec v12.5.0 \citep{arnaud96}. All quoted errors correspond to a 90 per cent confidence level, and energies are given in the observers frame (unless otherwise stated).

\begin{figure}
\includegraphics[width=\columnwidth]{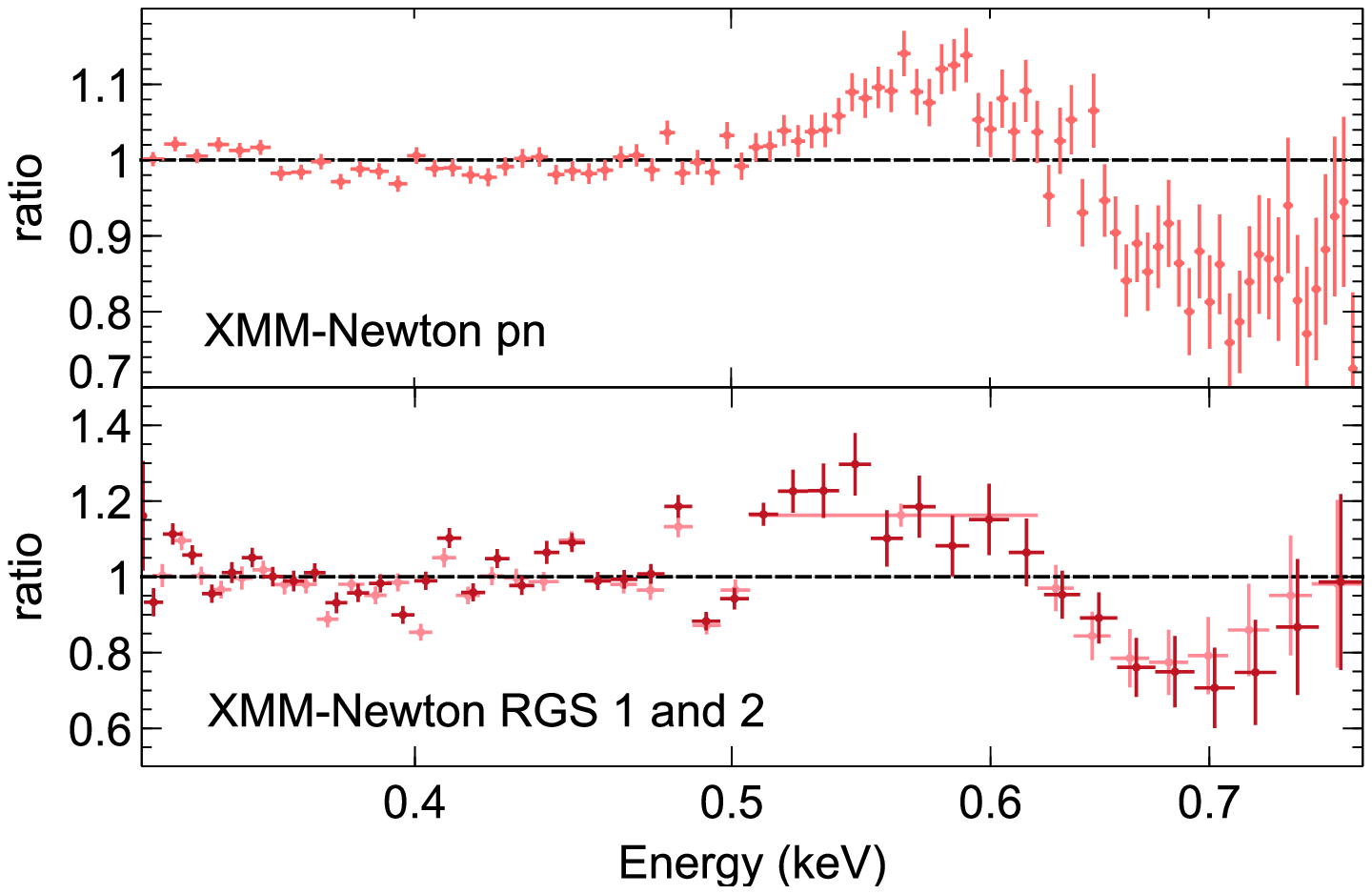}\\
\caption{The ratio of the early-time pn (top) and RGS (bottom) spectra to a single {\sc diskbb} model. The P~Cygni-esque feature is seen in both CCD and grating spectra, which demonstrates that this feature is not due to any residual pile-up in the CCD spectra. The RGS spectra shown in the plot are binned just for visual purposes.}
\label{pn_rgs}
\end{figure}

\section{Results}
\label{results}

Similar to several authors \citep{miller15,holoien16,vanvelzen16,brown17}, we begin by fitting all spectra with a $\sim 10^5$~K disk blackbody with neutral absorption. In Fig.~\ref{spec}, we show the combined early-time spectrum and combined late-time spectrum fit to {\sc phabs*diskbb} in {\sc XSPEC}. The ratio plot on the bottom panel shows that both early and late time spectra diverge from a simple disk blackbody at high energies. This hard tail can be well described by an additional Comptonization component or a higher temperature disk blackbody. Adding an additional disk blackbody improves the fit by $\Delta \chi^{2} = 50$ for two additional degrees of freedom for the early-time spectrum, and $\Delta \chi^{2} = 78$ for the late-time spectrum. Both the soft disk blackbody and the hard component decay over time, but the soft disk blackbody decays more quickly. In other words, the relative contribution of the hard tail increases over time. The flux ratio between the hard tail and the main soft thermal component for the early-time is 0.08\%, compared to 2.2\% at late times.


The early time spectra show a distinctive residual at around 0.6-0.8~keV. This can be seen clearly in Fig.~\ref{pn_rgs} in both the pn and RGS spectra. These panels show the ratio of the data to a disk blackbody model. We focus on the 0.3-0.8~keV energy range (where the RGS source spectrum dominates over the background). In the RGS spectrum (binned for visual purposes), one can see the low-energy features corresponding to the narrow low-velocity features found in \citet{miller15}, but we focus on the higher energy broad feature at 0.6-0.8~keV. The fact that this feature is evident in both pn and RGS spectra gives us confidence that this feature is not a pile-up artifact (see Appendix for more details). If we fit this feature with a simple phenomenological edge, we find $E_{\mathrm{zedge}}= 0.65 \pm 0.1$~keV in the source rest frame and $\tau=0.70\pm0.07$. 

\begin{figure}
\includegraphics[width=\columnwidth]{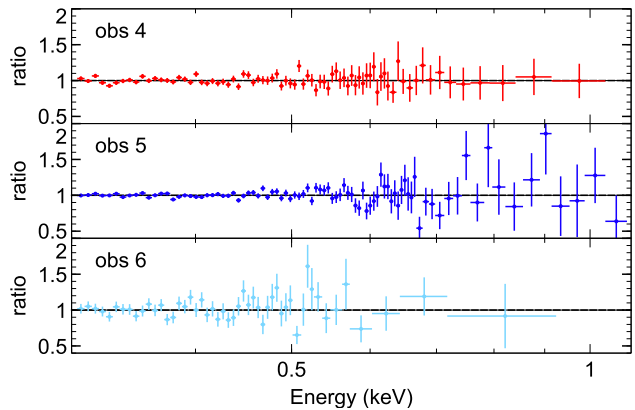}\\
\caption{The ratio of individual late-time spectra to a two component continuum model, consisting of two blackbody models. All late-time observations are well-fit by the continuum, and do not show the broad atomic feature at 0.6-0.8~keV that is present in the early-time spectra. }
\label{obs456}
\end{figure}

Interestingly, the feature at 0.6-0.8~keV is has disappeared or at least significantly decreased during the late-time spectra. The late-time spectra are well-described by the simple two disk blackbody continuum components with no additional absorption (Fig.~\ref{obs456}; Table~\ref{spec_table}). Including an additional absorption feature at 0.65~keV (the energy of the edge in the early-time spectra) only improves the fit by $\Delta \chi^{2} = 1.5$ for 1 degree of freedom (e.g. preferred at 80\% confidence). With this model, we place a 90\% upper limit on the optical depth of $\tau < 0.55$.




\begin{figure*}
\begin{center}
\includegraphics[width=0.85\textwidth]{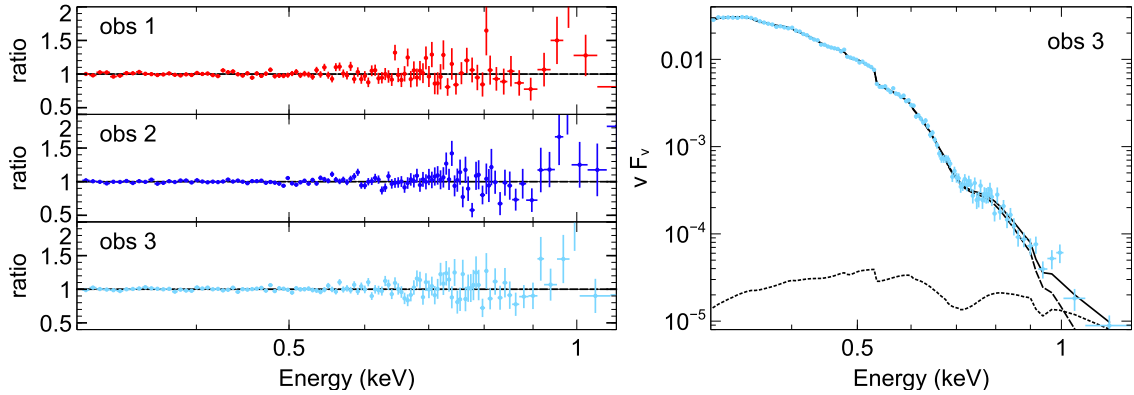}\\
\caption{{\em Left:} The ratio of individual early-time observations to their best fit continuum 
+ absorption from an UFO of 0.2c. {\em Right:} The unfolded spectrum and best-fit model
 of the third observation. Observations 1 and 2 show similar spectra. The black solid line is the total model; black dashed line is the absorbed, cooler blackbody (diskbb$_1$); black dotted is the absorbed, hotter blackbody (diskbb$_2$). }
\label{obs123}
\end{center}
\end{figure*}

\begin{table*}
\centering
\begin{tabular}{c c c c c c c c}
\hline
{\bf Model} & {\bf Parameter} & {\bf 1} & {\bf 2} & {\bf 3} & {\bf 4} & {\bf 5} & {\bf 6} \\
\hline
{\sc xstar} & nH ($10^{21}$~cm$^{-2}$) & $1.3\pm0.3$ & $1.2\pm 0.2$ & $1.8 \pm 0.3$ & -- & -- & -- \\
 & log($\xi$) log(erg~cm~s$^{-1}$) & $3.12 \pm 0.03$ & $3.11\pm 0.02$ & $3.16 \pm 0.05$ & -- & -- & --\\
 & $v_{\mathrm{out}} ($v/c$)$ & $0.23 \pm 0.02$ & $0.22 \pm 0.01$ & $0.22 \pm 0.01$ & -- &-- & --\\
diskbb$_1$ & $kT_{1}$ (eV) & $57\pm 2$ & $56.4 \pm 0.1$ & $56.8 \pm 0.1$ & $44.9 \pm 0.1$ & $41.6\pm 0.1$ & $38.0\pm 0.4$\\
diskbb$_2$ & $kT_{2}$ (eV)& $140^{+140}_{-48}$ & $195^{+135}_{-55}$ & $158^{+190}_{-58}$ & $213^{+110}_{-10}$ & $170^{+1400}_{-140}$ & $100^{+500}_{-50}$ \\
\hline
 & $\chi^{2}/{\mathrm{d.o.f.}}$ & $101/95$ & $137/107$ & 106/104 & 64/72 & 93/83& 50/45\\
\hline
\end{tabular}
\caption{Best fit model parameters for each observation. All observations require a `hard' tail (modelled as an additional diskbb component; `diskbb$_{2}$'), and only the first three observations require additional ionised absorption.}
\label{spec_table}
\end{table*}

The feature in the early-time spectra around 0.6.-0.8~keV seen in Fig.~\ref{pn_rgs} resembles a P~Cygni profile, and we explore this possibility through photoionization modelling using the {\sc XSTAR} code \citep{kallman01}. The default {\sc XSTAR} includes only thermal emission (i.e. electron impact excitation and recombination).  
We use a version of {\sc XSTAR} that includes resonance scattering, which is important for the emission component of P~Cygni profiles. {\sc XSTAR} produces the emission and absorption spectra for a spherical shell of gas irradiated by a continuum source. We produced a grid of photoionization models for varying column density and ionization parameter, assuming a $1.5\times10^{5}$~K blackbody continuum with a luminosity of $10^{45}$~erg~s$^{-1}$ (the unabsorbed luminosity from 1-1000~Ryd, (equivalent to 0.0136-13.6~keV), irradiating a constant density shell of $10^{10}$~cm$^{-3}$. We did build a grid including irradiation by the additional hard component, but this did little to change the overall absorption and emission features. The absorption feature is very broad (equivalent width of $50\pm20$~eV, corresponding to a FWHM velocity broadening of 0.2c) and so we produced {\sc XSTAR} grids with the maximum velocity broadening of 30,000~km/s. All abundances were fixed to solar values.

One of the most common and prominent lines in the spectra of AGN is the OVIII line at 0.653~keV. OVIII absorption provides the best fit to the pn spectra of ASASSN-14li at early times. The left panel of Fig.~\ref{obs123} shows the ratio of the first three spectra to the best fit models, which consist of two disk blackbody components absorbed through an outflow of 0.2c ({\sc phabs*zphabs*xstar*(diskbb$_1$+diskbb$_2$)} in {\sc XSPEC}; see details in Table~\ref{spec_table}). As an example of the complete model, the unfolded spectrum of the third observation is shown on the right panel of Fig.~\ref{obs123}.


While formally the absorbed blackbody models provide good fits to the data, one can see that there are systematic residuals in all three observations (most prominently, a peak at $\sim 1$~keV). We attempt to improve the model by including the corresponding emission component from the remainder of the spherical shell, not along our line of sight, to model the 1~keV residual as emission from Neon X at 1.08~keV. We allow for the outflow velocity and  normalization to be free, but fix the column density and ionization parameter to the absorption component. The additional component improves the fits slightly ($\Delta \chi^{2} = 2, 16, 4 $ for two additional degrees of freedom). The best-fit velocity of the emission component centers around the systemic velocity of the host, consistent with a typical emission component for a spherical outflow. However, the normalization is 10-100 times larger than the emission from a spherical shell geometry. The inclusion of the emission component (required only at the 1--3$\sigma$ level) suggests that the spherical symmetry of the reprocessor must break. This could be accomplished geometrically (i.e. perhaps the wind is an equatorial wind, where the solid angle of the emitting region is much larger than the absorption region) or by the absorbing gas having a slightly lower optical depth than the emitting region.

\section{Discussion and Conclusions}
\label{discuss}
In summary:\begin{enumerate}
\item Our analysis reveals an absorption feature at 0.6-0.7~keV in both CCD detectors and reflection grating spectrometers, which gives us confidence that this is not an instrumental effect.
\item The feature is evident only at early times, and is not seen in the observations taken $\sim 1$~year after the peak of the emission.
\item The early-time spectra are consistent with OVIII absorption through an ionised outflow of 0.2c.
\end{enumerate}

Super-Eddington accretion can lead to fast, radiation-driven outflows, which are commonly seen in MHD simulations of such systems (e.g. \citealt{coughlin14}, Dai et al., {\em in prep.}). Our analysis shows that the unabsorbed luminosity from 1-1000 Ryd (0.0136-13.6~keV) at early times is $\sim10^{45}$ erg~s$^{-1}$, which is $\sim L_{\mathrm{Edd}}$ (assuming a $10^{6.5}~M_{\odot}$ black hole; \citealt{vanvelzen16}), however lower bolometric luminosities have also been suggested \citep{holoien16}. Recent work by Dannen, Proga \& Kallman, {\em in prep}, suggest that high-ionization outflows can still be produced via UV line driving in sub-Eddington systems. 

The wind is not observed (or at least is significantly diminished) at late-times ($\sim1$~year after the discovery), which is consistent with classical TDE fallback theory. One year after mass fallback peak, the fallback rate is $\sim 1\dot{M}_{\mathrm{Edd}}$. If we observed a super-Eddington radiatively-driven outflow, it will have `turned-off' 1~year after the event, and could defuse out into the ambient medium, thus becoming unobservable. Alternatively, it is possible that the wind still exists, but no longer intercepts our line of sight.  

The minimum launching radius of the wind can be estimated from the radius at which the observed velocity corresponds to the escape velocity, $r=2GM_{{\mathrm{BH}}}/c^{2}$. This corresponds to a radius of $\sim 38~r_{\mathrm{g}}$. 

We can use this radius to estimate the mass outflow rate of the wind, following the relation in \citet{nardini15}: $\dot{M}_{\mathrm{out}} = \Omega m_{\mathrm{p}} N_{\mathrm{H}} v_{\mathrm{out}} r ,$ where $\Omega$ is the solid angle of the wind, which, for simplicity, we take to be $2\pi$. The minimum mass outflow rate (using the minimum radius) is then $\sim 1 \times 10^{21}$~g~s$^{-1}$ ($\sim 0.0002~\dot{M}_{\mathrm{Edd}}$).

In most studies of winds in Seyfert galaxies, the maximum mass outflow rate is estimated using the definition of the ionisation parameter and by assuming that the thickness of the absorber does not exceed its distance to the black hole, such that $r_{\mathrm{max}}\equiv L_{\mathrm{ion}}/\xi N_{\mathrm{H}}$. With our parameters, this leads to a maximum outflow rate of $\sim10^{29}$~g~s$^{-1}$.
However, we can get tighter constraints on the maximum outflow rate from the natural assumption that the total outflow mass cannot exceed the amount of stellar debris falling back towards the black hole. We assume the fallback follows the light curve and deduce that the $M_{\mathrm{fallback}} \sim (t-t_{0})^{-5/3}$ with $t_0 \sim 56950$~MJD \citep{miller15, vanvelzen16}. Therefore, the three early observations happen within $\sim70$~days since $t_0$, and the total mass that falls back during this period is $\sim0.1~M_{\odot}$. Assuming 10\% of the fallback material becomes outflow \citep{sadowski16, mckinney15}, then the maximum outflow rate is $3 \times 10^{24}$~g~s$^{-1}$ ($\sim0.8\dot{M}_{\mathrm{Edd}}$), and the corresponding maximum wind launching radius is $\sim 8\times 10^{4}~r_{\mathrm{g}}$.

The X-rays from ASSASN-14li are also absorbed by a low-velocity wind of $\sim 300$~km~s$^{-1}$, which were interpreted as absorption through a super-Eddington wind or through a filament of stellar debris \citep{miller15}. The low velocity suggests that this wind is produced at larger radii than the fast wind reported here. The low and high velocity winds have similarly high ionization parameters, suggesting that the density of the low-velocity absorber is lower than that of the high-velocity absorber. Perhaps the fast outflow slows down as it collides with the debris stream or some other medium and produces a lower-velocity component. If this gas diffuses out, it could cause the low-velocity, high-ionization absorption lines.

The mass and velocity of the outflow estimated from these X-ray observations are very similar to those derived from the radio observations presented in \citet{alexander16}. These authors interpreted the radio emission as an outflow of mass $3-70\times 10^{-5} M_{\odot}$, outflowing at a velocity of $0.04-0.12$c. That said, simultaneous fast winds and relativistic jets are common to MHD simulations of super-Eddington flows, and so our observations are also consistent with the interpretation of the radio emission as a collimated jet (e.g. \citealt{vanvelzen16,pasham17}).



\section*{Acknowledgements}

Thank you to the anonymous reviewer for helpful comments. EK thanks {\em XMM-Newton} PI, Norbert Shartel, and Matteo Guinazzi for consulting on the data reduction for ASASSN-14li, and Nathan Roth, Cole Miller for interesting discussions on TDEs and Daniel Proga and Randall Dannen for discussions on UV line driving in ionised winds. EK thanks the Hubble Fellowship Program. Support for Program number HST-HF2-51360.001-A was provided by NASA through a Hubble Fellowship grant from the Space Telescope Science Institute, which is operated by the Association of Universities for Research in Astronomy, Incorporated, under NASA contract NAS5-26555.



\bibliographystyle{mnras}
\bibliography{tde} 



\appendix 
\renewcommand{\thefigure}{A\arabic{figure}}
\setcounter{figure}{0}

\begin{figure}
\begin{center}
\includegraphics[width=\linewidth]{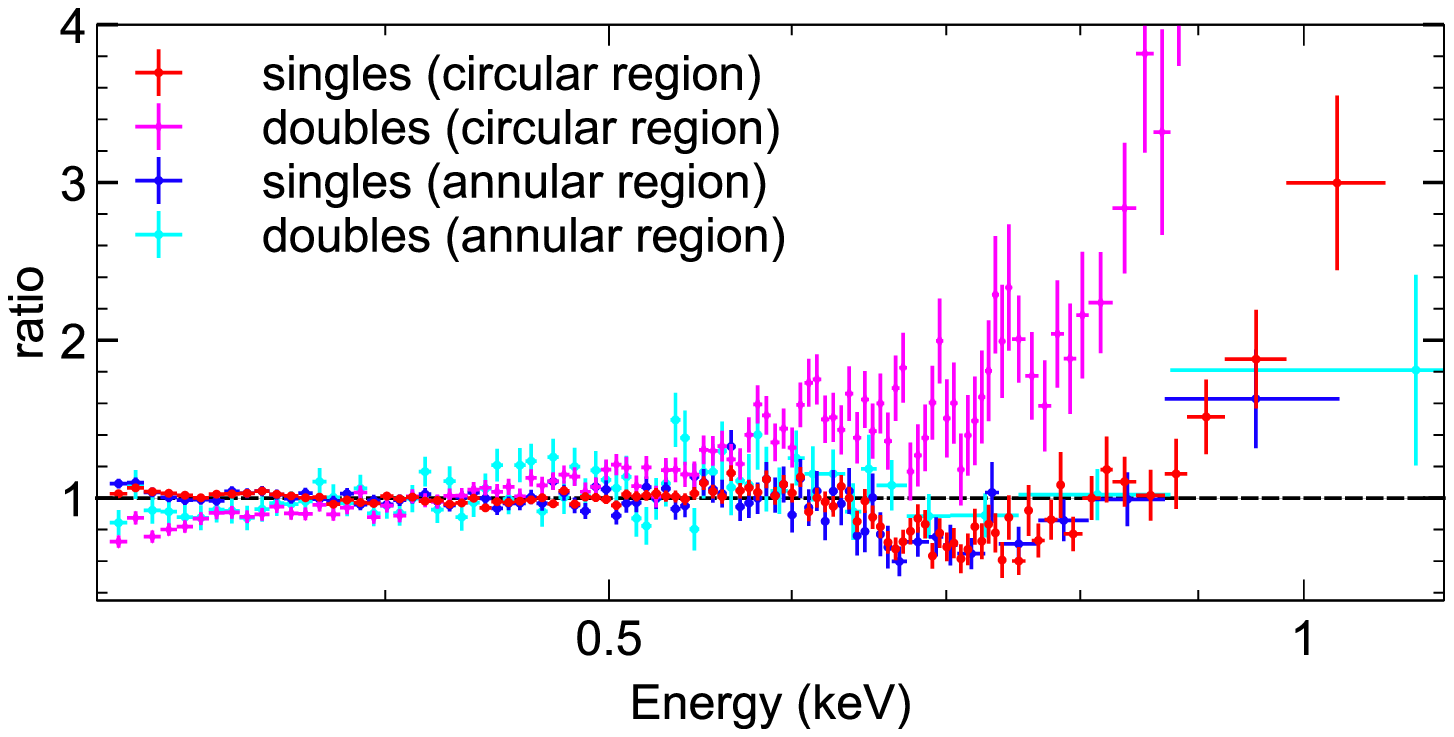}\\
\label{appendix}
\caption{The ratio plot demonstrating the effects of pile-up on a full circular region (red colors) compared to an annular region (blue colors). The fact that the single and double events spectra for the annular source region look nearly identical gives us confidence that we have reduced pile-up in the pn spectrum.}
\end{center}
\end{figure}

\section*{Appendix: Handling pile-up}
Pile-up is an unfortunate and inevitable consequence of the finite read-out time of CCD detectors. Pile-up was an issue in this very bright, soft X-ray spectrum, and in this appendix, we show the steps taken to minimize these effects.

Pile-up can be seen most clearly by comparing the single events spectrum and double events spectra from the full 35 arcsecond circular source region. Fig.~A1 shows an example using the third observation, which had minimal background flaring. This figure shows the ratio of the single and and double event spectra (in red and magenta, respectively) fit to a simple blackbody model with neutral absorption (only normalization allowed to vary).  The double events spectra are more susceptible to pile-up, and one can see clearly that the magenta double events spectrum is much harder than the red single events spectrum. Because the single and double events spectra look markedly different, we know that pile-up is likely affecting at least the double events spectrum (if not both).

The most common way to minimize pile-up is to excise the central pixels of the circular region where pile-up is the worst. As a first check, we used the {\em XMM-Newton} SAS tool, {\sc epatplot}, to establish how large the excision region needed to be such that the number of single and double events were within the expected values (see Table~\ref{obs_table} for details). Fig.~A1 shows the single and double events spectra for an annular source region (in blue and cyan, respectively). The double events spectrum (cyan) no longer has the hard tail, which indicates that much of the pile-up has been removed. 
The double events spectrum looks the same as the single events spectrum to within errors. There is a slight depression of both double events spectra below 0.35 keV, which is likely a calibration error due to the higher energy electronic noise component in double events spectra (Matteo Guinazzi, private communication). The fact that single and double events spectra look the same when excising a central region gives us confidence that this technique is cleaning our spectrum of pile-up.

Excising a large portion of the central pixels comes at the cost of losing many of the counts in the spectrum, thus decreasing the overall signal-to-noise. We found that the singles spectrum for the full 35 arcsecond circular region (red) looked identical to our `cleanest' spectrum, i.e. the single events spectrum using the annular region (blue). We therefore opted to use the singles spectrum for the entire circular region to improve signal-to-noise.

We emphasize that the hard excess found in all spectra (even in our pile-up cleaned spectra) appears 
to increase in the late-time observations (when the overall countrate has decreased). If the hard tail was merely an artifact due to pile-up, we would expect the hard tail to decrease with decreasing luminosity, while, in fact, the opposite trend is found.  Furthermore, it is unlikely that the $\sim 0.6-0.8$~keV feature is an effect of residual pile-up because the signature is also seen in the RGS spectra, where the source image is dispersed across the chip. Lastly, the $\sim 0.6-0.8$~keV feature is seen in both EPIC-MOS detectors. While this is nice confirmation, the MOS detectors are even more subject to pile-up than the pn.


\bsp	
\label{lastpage}
\end{document}